\documentstyle[sprocl,epsfig,amssymb]{article}

\def\be{\begin{equation}}
\def\ee{\end{equation}}
\def\bea{\begin{eqnarray}}
\def\eea{\end{eqnarray}}

\begin{document}

\vspace*{-1.3cm}

\begin{flushright}
Edinburgh 2004/20\\[0.5cm]
\end{flushright}

\title{The Higgs and Neutralino Sectors of the Next--to--Minimal Supersymmetric Standard Model}

\author{D.~J.~Miller}

\address{
Department of Physics and Astronomy, University of Glasgow,\\ Glasgow G12 8QQ, Scotland \\ {\it and}\\
School of Physics, University of Edinburgh, Edinburgh EH9 3JZ, Scotland}


\maketitle\abstracts{ The Next--to--Minimal Supersymmetric Standard
Model (NMSSM) includes a Higgs iso-singlet superfield in addition to
the two Higgs doublet superfields of the minimal supersymmetric
extension.  The Higgs sector and neutralino sectors of this model are
examined within the context of a future $e^+e^-$ linear collider.}
  
\vspace*{-1.4mm}

\section{Introduction} \vspace*{-1.4mm}

One of the main tasks for the next generation of colliders will be to
determine the mechanism of electroweak symmetry breaking. If this
turns out to be the Higgs mechanism together with supersymmetry, it
will then become essential to distinguish the minimal supersymmetric
model from its non-minimal extensions. In this talk, I will discuss
one such extension: the Next--to--Minimal Supersymmetric Standard
Model~\cite{Miller:2003ay,Choi:2004zx} (NMSSM) in which an iso--singlet
Higgs superfield $\hat{S}$ is introduced in addition to the two Higgs
doublets superfields of minimal supersymmetry, $\hat{H}_{u,d}$. Such
an extension offers a possible solution of the $\mu$ problem,
generating in a natural way, a value of the order of the electroweak
breaking scale $v$; this is achieved by identifying $\mu$, apart from
the ${\cal O}(1)$ coupling, with the vacuum expectation value of the
scalar component $S$ of the new iso--singlet field.

The superpotential of the NMSSM is given by
\begin{eqnarray}
W=W_Y +\lambda \hat{S}(\hat{H}_u \hat{H}_d)+\frac{1}{3}\kappa\hat{S}^3
\label{eq:superpotential}
\end{eqnarray}
where $W_Y$ denotes the usual MSSM Yukawa components. The first
additional term regenerates the $\mu$-term of the MSSM when $S$ gains
a vacuum expectation value, while the second provides an explicit
breaking of a $U(1)$ Peccei-Quinn symmetry which would otherwise be
present; if not explicitly broken this additional symmetry would lead
to a (near) massless pseudoscalar Higgs boson during spontaneous
symmetry breaking.  The two parameters $\lambda$ and $\kappa$ are
dimensionless, and are bounded by $\lambda, \kappa \lesssim 0.7$ at
the electroweak scale if they are to remain weakly interacting up to
the GUT scale. Also, renormalisation group running favours values of
$\kappa$ lower than $\lambda$ at the electroweak scale.

\vspace*{-1.4mm}

\section{The Higgs Sector} \vspace*{-1.4mm}

The complex superfield $\hat S$ gives rise to two extra Higgs bosons,
one scalar and one pseudoscalar, enlarging the Higgs
sector~\cite{Miller:2003ay} to three neutral scalar fields, two
neutral pseudoscalar fields and two charged fields. The Higgs
potential will also contain soft supersymmetry breaking terms,
parameterized by soft masses and the soft trilinear parameters
$A_{\lambda}$ and $A_{\kappa}$ corresponding to the new terms in the
superpotential. The vacuum minimization conditions allow one to
replace the soft masses with the electroweak scale $v=246$~GeV, and
two ratios of vacuum expectation values, $\tan \beta \equiv \langle
H_u \rangle / \langle H_d \rangle$ and $\tan \beta_s \equiv \sqrt{2}
\langle S \rangle /v$.  It is useful to replace the parameter
$A_{\lambda}$ with the mass of the heaviest pseudoscalar Higgs
$M_{A_2}$, in analogy to the usual procedure adopted in the MSSM.

\begin{figure}[ht]
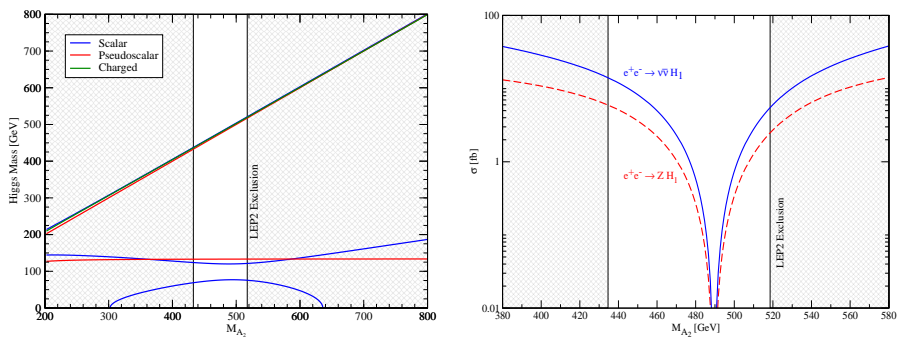
 
     \begin{center}
     \vspace*{.2cm}
     \begin{tabular}{cc}
     \mbox{\epsfig{file=masses_higgs.eps,width=5.7cm,clip=true}}&
     \mbox{\epsfig{file=tesla1_500.eps,width=5.7cm,height=4.35cm,clip=true}}
     \end{tabular}
     \end{center}
     \caption{The one-loop Higgs boson masses {\it (left)} and some
              tree-level production cross-sections for the lightest
              Higgs boson at a $\sqrt{s}=500$~GeV $e^+e^-$ linear
              collider {\it (right)}, as a function of $M_{A_2}$, for
              $\lambda=0.3$, $\kappa=0.1$, $A_{\kappa}=-100$~GeV,
              $\tan \beta=\tan \beta_s = 3$ and a running top mass of
              $m_t=165$~GeV, and maximal mixing.}
     \label{fig:higgs}
     \end{figure}

The one-loop Higgs mass spectrum as a function of the heavy
pseudoscalar mass $M_{A_2}$ is shown in Fig.\ref{fig:higgs} {\it
(left)} for a scenario with a low value of $\kappa$.
The shaded region is excluded by LEP2, restricting the allowed heavy
pseudoscalar mass to values near $M_{A_2} \approx \mu \tan
\beta$. Notice that rather light Higgs bosons $\sim 70$~GeV are still
not ruled out~\cite{Miller:2004uh}, due to a reduced coupling of the
Higgs boson to gauge bosons which lowers the Higgs-strahlung
cross-section. Such a Higgs boson would be very difficult to see at
the LHC; its decay is mainly hadronic $H_1 \to b \bar b$ ($H_1 \to
\gamma \gamma$ is also suppressed over most of the region) and will be
swamped by huge QCD backgrounds. The lightest Higgs boson production
cross-sections at a $\sqrt{s}=500$~GeV $e^+e^-$ linear collider are
shown in Fig.\ref{fig:higgs} {\it (right)}, where one can see that the
coupling to gauge bosons switches off entirely at $M_{A_2} \approx
490$~GeV. Although this results in a small range of $M_{A_1}$ still
being inaccessible, these search channels cover most of the allowed
parameter space. As the centre-of-mass energy of the collider is
increased these cross-sections scale in the usual way.

The Higgs bosons comprised predominantly of the singlet fields
increase in mass as $\kappa$ is increased, with approximate squared
masses for the scalar and pseudoscalar given by $\kappa \langle S
\rangle \left( \kappa \langle S \rangle +A_{\kappa} \right)$ and
$-3\kappa \langle S \rangle A_{\kappa}$ respectively. Once the scalar
is heavy enough it may decay $ZZ^*$ and should be clearly visible at
the LHC for values of $M_{A_2}$ away from the $Z$ coupling switch-off
point.

\vspace*{-1.4mm}

\section{The Neutralino Sector}\vspace*{-1.4mm}

\begin{figure}[ht]
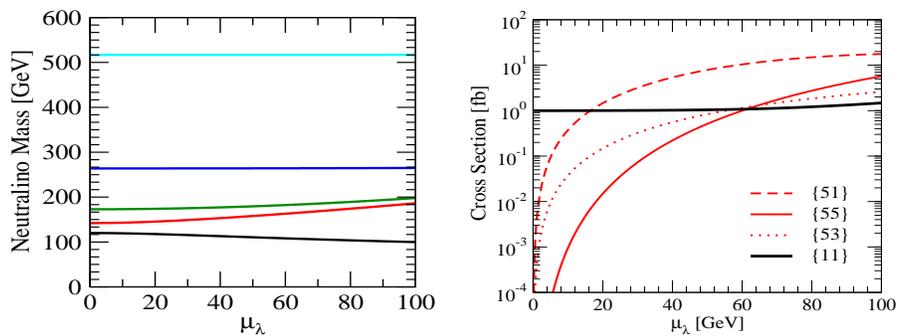

     \begin{center}
     \begin{tabular}{cc}
     \mbox{\epsfig{file=nmass_ml.eps,width=5.7cm,height=4.35cm,clip=true}}&
     \mbox{\epsfig{file=prod.eps,width=5.7cm,height=4.35cm,clip=true}}
     \end{tabular}
     \end{center}
     \caption{The neutralino mass spectrum {\it (left)} and the
              cross-sections for $e^+e^- \to \tilde \chi_i^0 \tilde
              \chi_j^0$ at a $\sqrt{s}=500$~GeV $e^+e^-$ linear
              collider {\it (right)}, as a function of
              $\mu_{\lambda}$, for $\mu_{\kappa}=120$~GeV,
              $\mu=170$~GeV, $\tan \beta=3$, $M_1=250$~GeV and
              $M_2=500$~GeV.}
     \label{fig:neut}
     \end{figure}

In contrast to the Higgs sector, the neutralino
sector~\cite{Choi:2004zx} is complemented only by the familiar SU(2)
and U(1) gaugino mass terms, resulting in a much less complex
parameter space. The extra singlet superfield adds an extra higgsino
to the spectrum, often called a {\it singlino}, resulting in five
neutralino states. We denote the singlino dominated neutralino $\tilde
\chi_5^0$, with $\tilde \chi_1^0$--$\tilde \chi_4^0$ denoting the
other four neutralinos in order of ascending mass. The neutralino
spectrum for an example scenario is shown in Fig.\ref{fig:neut} {\it
(left)} as a function of $\mu_{\lambda} \equiv \lambda v/\sqrt{2}$.
In this scenario, the singlino dominated neutralino {\it (black)} is
the lightest neutralino (and the LSP) with a mass of approximately
$\mu_{\kappa} \equiv 2 \kappa \langle S \rangle$. We have chosen to
fix $\mu$ in this section rather than $\tan \beta_s$ as previously
because of the importance of $\mu$ to the higgsino spectrum. If the
singlino is the LSP (as here) it will be copiously produced at the LHC
in squark and gluino cascade decays. A very decoupled state can give
rise to macroscopic flight distances for both the decays $\tilde
\chi_1^0 \to \tilde \chi_5^0 l^+l^-$ and $\tilde l_R \to \tilde
\chi_5^0 l$, but this requires rather low values of $\lambda$, with
$\mu_{\lambda}=1$~GeV producing a flight distance of order a $\mu m$
and order a $nm$ for the two respective decays. Also shown in
Fig.\ref{fig:neut} {\it (right)} are the cross-sections for $e^+e^-
\to \tilde \chi_i^0 \tilde \chi_j^0$ at a linear collider, for
production of singlino-like ($\tilde \chi_5^0$), gaugino-like ($\tilde
\chi_1^0$) and higgsino-like ($\tilde \chi_3^0$) neutralinos. With the
anticipated integrated luminosity of $\int {\cal L} =1 \; {\rm
ab}^{-1}$, sufficiently large event rates of order $10^3$ are expected
if $\mu_{\lambda}$ is not too small.

Comparing the characteristic higgsino mass-scale, $\mu = \lambda
\langle S \rangle$, with the characteristic singlino mass-scale,
$\mu_{\kappa} = 2\kappa \langle S \rangle$, we see that $\kappa$ does
not need to become too large, $\kappa \gtrsim \lambda/2$, for
$\tilde\chi_5^0$ to become heavier than $\tilde\chi_1^0$. In this
case, the singlino will no longer be the LSP and will decay to $\tilde
\chi_1^0$. Such a neutralino sector would be very difficult to
distinguish from that of minimal supersymmetry.

\vspace*{-1.4mm}

\section{Conclusions} \vspace*{-1.4mm}

The NMSSM is a viable extension to minimal supersymmetry, with an
in-built solution to the $\mu$-problem. In addition to the usual Higgs
and neutralino states, it provides an extra scalar and pseudoscalar
Higgs field, and an extra neutralino. All of these extra fields
increase in mass as $\kappa$ (the parameter quantifying explicit
Peccei-Quinn symmetry breaking) is increased. For small values of
$\kappa$, the extra scalar Higgs field will be difficult to see at the
LHC, while for medium to large values of $\kappa$, the extra
neutralino will be difficult to identify. Although for much of the
allowed parameter space, either the Higgs sector or the neutralino
sector will display manifestly non-minimal structure, there is a large
part of parameter space were such a non-minimal structure could be
hidden. There are also small regions of parameter space where the
coupling of the extra scalar to gauge bosons switches off entirely. It
will therefore be extremely important to examine the Higgs and
neutralino sectors in a precision environment, such as an $e^+e^-$
linear collider.\\

\noindent {\bf Acknowledgements:}
DJM would like to thank S.~Y.~Choi, R.~Nevzorov, S.~Moretti and
P.~M.~Zerwas for the fruitful collaborations in which this research
was carried out.

\end{document}